\documentstyle[fleqn,espcrc2,epsf]{article}

\def\beq{\begin{equation}}
\def\eeq{\end{equation}}
\def\bea{\begin{eqnarray}}
\def\eea{\end{eqnarray}}

\def\np#1#2#3{Nucl.\ Phys.\ B#1 (19#3) #2}

\def\pr#1#2#3{Phys.\ Rev.\ D #1 (19#3) #2}

\def\mafigura#1#2#3#4{
  \begin{figure}[hbtp]
    \begin{center}
      \epsfxsize=#1
      \leavevmode
      \epsffile{#2}
    \end{center}
    \caption{#3}
    \label{#4}
  \end{figure} }

\newcommand{\eq}[1]{eq.~(\ref{#1})}
\newcommand{\as}{\alpha_s}

\newcommand{\yc}{y_{c}}
\newcommand{\rb}{r_b}
\newcommand{\el}[1]{^{(#1)}}
\newcommand{\non}{\nonumber}

\begin{document}

\renewcommand{\thefootnote}{\fnsymbol{footnote}}
\title{The running of the $b$-quark mass from LEP data
\thanks{Supported in part by CICYT, Spain, 
under grant AEN-96/1718 and in part by DFG Projekt No. Ku 502/8-1,
Germany.}}
\author{
Germ\'an Rodrigo~\address{
Intitut f\"ur Teoretische Teilchenphysik, Universit\"at
Karlsruhe, 76131 Karlsruhe, Germany}\thanks{
Supported in part by CSIC-Fundaci\'o Bancaixa.
On leave from Departament de F\'{\i}sica Te\`orica, IFIC,
CSIC-Universitat de Val\`encia, 46100 Burjassot, Val\`encia, Spain.},
Arcadi Santamaria~\address{
Departament de F\'{\i}sica Te\`orica, IFIC,
CSIC-Universitat de Val\`encia, 46100 Burjassot, Val\`encia, Spain} 
and
Mikhail Bilenky~\address{
Institute of Physics, AS CR, 18040 Prague 8, and 
Nuclear Physics Institute, AS CR, 25068 \v{R}e\v{z}(Prague),
Czech Republic}\thanks{On leave from JINR, 141980 
Dubna, Russian Federation.} }

\begin{abstract}
Next-to-leading order QCD corrections
to three jet heavy quark production in $e^+e^-$ collisions,
including quark mass effects,
are presented. The extraction of the $b$-quark mass from LEP
data is considered and the first experimental evidence for
the running of a quark mass is discussed.
\end{abstract}

\maketitle

\section{Introduction}

The question of the origin of the masses of quarks and leptons
is one of the unresolved 
puzzles in present high energy physics. To answer this question one needs 
to know precisely their value. However, quarks are not free and their mass
has to be interpreted more like a coupling than an inertial parameter
and it can run if measured at
different scales. Moreover, in the standard model (SM) all
fermion masses come from Yukawa couplings and those also run with 
the energy. To test fermion mass models one has to run masses 
extracted at quite different scales to the same scale and compare them
with the same "ruler". This way, for instance, one can check that in some
unified models the $b$-quark mass and the $\tau$-lepton mass,
although different
at threshold energies they could be equal at the unification scale.

However, the running of fermion masses, although predicted by
quantum field theory, has not been tested experimentally until now.
The reason being that for energies $\sqrt{Q^2}$ much higher than
the fermion mass, $m_q$, the mass effects become negligible since
usually they are suppressed by $m_q^2/Q^2$. 

While this argument is correct for total cross sections for production of
heavy quarks, it is not completely true for quantities that depend on other
variables.
In particular, it is not true for jet cross sections which depend
on a new variable, $\yc$ (the jet-resolution parameter
that defines the jet multiplicity) and
which introduces a new scale in the analysis, $E_c=\sqrt{Q^2\: \yc}$. Then,
for small values of $\yc$ there could be contributions coming like
$m_q^2/E_c^2 = (m_q^2/Q^2) /\yc$ which could enhance the mass effect
considerably. In addition mass effects could also be enhanced by
logarithms of the mass. For instance, the ratio of the phase space
for two massive quarks and a gluon to the phase
space for three massless
particles is $1+8 (m_q^2/Q^2) \log(m_q/Q)$.
At $Q^2=m_Z^2$ and for the bottom quark this gives a 7\% effect,  
for $m_b=5$~GeV and a 3\% effect for $m_b=3$~GeV.

The high precision achieved at LEP makes heavy quark mass
effects relevant.
In fact, they have to be taken into account in the
tests of the flavour independence of $\as(m_Z)$~\cite{chrin,l3}. 
This in turn means that mass effects have already been seen. 
One can reverse the question
and ask about the possibility of measuring the mass of the bottom
quark, $m_b$, at LEP by assuming the flavour universality of the strong
interactions. 

In~\cite{bilenky.rodrigo.ea:95} we showed that mass effects
in three-jet production at LEP are large enough
to be measured.
The observable proposed as a means to extract
the bottom-quark mass from LEP data was the 
ratio~\cite{chrin,bilenky.rodrigo.ea:95}
\begin{equation}
R^{bd}_3 \equiv \frac{\Gamma^b_{3j}(\yc)/\Gamma^b}
{\Gamma^d_{3j}(\yc)/\Gamma^d}~.
\label{eq:r3bd_def}
\end{equation}
In this equation $\Gamma^q_{3j}(\yc)/\Gamma^q$ is the three-jet fraction
and $q$ denotes the quark flavor.
In this ratio and at the leading order (LO) the quark mass effects 
can be as large as 1\% to 6\%,
depending on the value of the mass and on 
the jet-resolution parameter, $\yc$.

The three jet decay width is defined by the jet-clustering
algorithms (see e.g. \cite{bethke.kunszt.ea:92}). 
The effect of hadronization is, in principle, small and has been 
estimated using the
Montecarlo approach~\cite{fuster.cabrera.ea:96,marti.fuster.ea:97}.

Since the measurement of $R^{bd}_3$ is done
far away from the threshold of $b$-quark production it can be used
to test the running of a quark mass as predicted by QCD.

However, as we discussed in \cite{bilenky.rodrigo.ea:95},
the leading order calculation does not distinguish among the
different definitions of the quark mass: perturbative pole mass, $M_b$,
running mass at $M_b$-scale, or running mass at $m_Z$-scale. 
As the numerical difference is significant when the different
definitions of masses are used in LO calculations, 
in order to correctly take 
into account mass effects, it is necessary to 
perform a 
complete next--to--leading order (NLO) calculation of 
three-jet ratios including quark masses.

Although, heavy quark production has been considered in a large variety
of processes and, in particular, in $Z$-boson decays 
\cite{rev,djouadi.kuhn.ea:90,jersak.laermann.ea:82,ballestrero.maina.ea:92},
there are very few NLO calculations of heavy quark jet production
taking into account complete mass effects (
In \cite{beenakker.kuijf.ea:89} and \cite{laenen.riemersma.ea:93}
this was done for gluon-gluon fusion and virtual-photon production of 
heavy quarks).  
Only very recently NLO calculations of heavy quark jet production  
in $e^+ e^-$ collisions have become
available~\cite{rodrigo:96,rodrigo.santamaria.ea:97*a,rodrigo.santamaria.ea:97*b,bernreuther.brandenburg.ea:97,nason.oleari:97}.
Here we overview our calculation. Its final results 
were presented in 
\cite{rodrigo.santamaria.ea:97*a}
and have been used by the DELPHI Collaboration to extract the 
$b$-quark mass at the $m_Z$ 
scale~\cite{fuster.cabrera.ea:96,marti.fuster.ea:97}
showing clearly that indeed the $b$-quark mass runs from
$\mu=m_b$ to $\mu=m_Z$ as predicted by QCD.

\section{Jet ratios with heavy quarks at NLO}

The decay width of the $Z$-boson into three jets with a heavy 
quark can be written as follows 
\beq
\Gamma^{b}_{3j} = \frac{m_Z g^2 \alpha_s}{c_W^2 64 \pi^2}
\left[g_V^2 H_V(y_c,r_b) + g_A^2 H_A(y_c,r_b)\right]~,
\label{eq:gamma3jets}
\eeq
where $g$ is the SU(2) gauge coupling constant,
$c_W$ and $s_W$
are the cosine and the sine of the weak mixing angle,
$g_V =-1+4/3 s_W^2$ and $g_A=1$ are the vector and axial-vector 
coupling of the 
$Z$-boson to the bottom quark
and $\alpha_s$ is the strong coupling constant.
Functions $H_{V(A)}(\yc,\rb)$ 
contain all the dependences on $\yc$ and the quark
mass, $\rb = (M_b/m_Z)^2$, for the different
algorithms. 
These functions can be expanded in $\as$ as
\bea
\label{eq:hexpansion}
& & H_{V(A)} = A^{(0)}(y_c) + \frac{\as}{\pi} A^{(1)}_{V(A)}(y_c) \\
& & + r_b \left[ B_{V(A)}^{(0)}(y_c,r_b) 
+ \frac{\as}{\pi} B_{V(A)}^{(1)}(y_c,r_b) \right]
+ \cdots~. \non
\eea
Here,  $A\el{0}$, is the tree-level contribution in the 
massless limit. It 
is the same function for the 
vector and the axial-vector
parts and it is known for the different jet-clustering algorithms
in analytic form.
The function $A^{(1)}_{V(A)}$ gives the QCD NLO correction for massless quark.
This function is also known
for the different jet-clustering algorithms
\cite{bethke.kunszt.ea:92}~\footnote{With our choice of
the normalization $A^{(0)}(y_c)=A(y_c)/2$ and 
$A^{(1)}_V(y_c)=B(y_c)/4$, where $A(y_c)$ and $B(y_c)$ are 
defined in~\cite{bethke.kunszt.ea:92}.}. 
Note that even in the chiral limit these corrections are different for 
the vector and axial-vector parts. However, the difference, which is due 
to the one-loop triangle diagrams~\cite{hagiwara.kuruma.ea:91} is rather
small. The net effect of these triangle diagrams is smaller than 
$10^{-4}$ in  $R_3^{bd}$.
In the ratio $R_3^{bl}$, which is similar to the one defined 
in~\eq{eq:r3bd_def} but
normalized to the sum of three light flavours, 
$l$=($u$,$d$,$s$),
the triangle anomaly produces a shift of $+2\cdot 10^{-3}$.
This has been taken into account in the experimental 
analysis~\cite{marti.fuster.ea:97}.
Functions $B_{V(A)}$ take into account 
residual mass effects,
once the leading dependence in $\rb$ has been factorized. 
The tree-level contributions, $B_{V(A)}\el{0}$,
were calculated numerically in \cite{bilenky.rodrigo.ea:95} for the 
different algorithms and results were presented in the form of 
simple fits to the numerical results.  
Finally, the functions $B_{V(A)}\el{1}$, 
contain the NLO corrections depending on the
quark mass.

Note that the way we write $H_{V(A)}$ in
\eq{eq:hexpansion} is not an expansion for small $\rb$. We keep
the exact dependence on $\rb$ in the functions  $B_{V(A)}$.
Factoring out $\rb$  makes it
easier to analyze the 
massless limit and the dependence of the results
on $\rb$ in the region of interest. This means that
our results can also be adapted, by including the photon exchange, to compute
the $e^+ e^-$-cross section into three jets outside the $Z$-peak at 
lower energies or at higher energies for top quark production.

At the NLO we have contributions to the three-jet cross section from three-
and four-parton final states. 
For diagrams with emission of 
four real quarks, that can mix different flavours, we take the
convention that the flavour is defined by the quark coupled 
directly to the Z-boson. Therefore, events with emission of 
a heavy quark pair radiated of a light $q\bar{q}$ are classified 
as light events despite the fact that $b$ quarks are present in
this four-fermion final state.
From the theoretical point of view this avoids the appearance of 
large logarithms on the quark mass. The same convention 
was considered in the experimental analysis~\cite{marti.fuster.ea:97}
therefore allowing for a consistent comparison.

One-loop three-parton
amplitudes are both infrared (IR)
and ultraviolet (UV) divergent.  Therefore, some regularization procedure 
is needed. We use
dimensional regularization 
for both IR and UV divergences, because
it preserves the QCD Ward identities. 
The three-parton transition amplitudes can be expressed in terms of a few
scalar one-loop integrals \cite{rodrigo.santamaria.ea:97*b}. 
The result contains poles in $\epsilon=(4-D)/2$,
where $D$ is the number of space-time dimensions. Some of the poles
come from UV divergences and the other come from IR divergences.
The UV divergences,
however, are removed after the renormalization of the parameters
of the QCD lagrangian.
After that we obtain analytical expressions, 
which contain terms proportional to the IR poles and
finite contributions. 

The four-parton transition amplitudes are also IR divergent.
These IR divergences cancel the corresponding IR poles
coming from the virtual corrections according to
\cite{bloch.nordsieck:37}. 
Two different methods of analytic cancelation of IR singularities
have been developed:
the {\it phase space slicing method}~\cite{slicing}
and the {\it subtraction
method}~\cite{bethke.kunszt.ea:92,ellis.ross.ea:81,subtraction}.
We follow the first approach.
The four-parton transition probabilities for 
$Z\rightarrow b\bar{b} gg (b\bar{b} q\bar{q})$ are split in two parts.
The first part contains the terms which are divergent when one gluon
is soft or two gluons (or light quarks) are collinear. These terms
are integrated analytically in arbitrary $D$ dimensions in the 
soft and collinear regions of phase space. This way we obtain
the IR singular contributions of four partons in the three-jet region
and show that they are canceled exactly by the tree-parton 
contribution.
The second part, corresponding to the radiation
of hard gluons, gives rise to finite contributions and can
be calculated in $D=4$ dimensions.
The three-jet $Z$-width is obtained by integrating both, 
renormalized three-parton contribution and four-parton transition
probabilities, in the three-jet phase-space region defined by the different 
jet-clustering algorithms. This quantity is
infrared finite and well defined.

Following Ellis, Ross and Terrano \cite{ellis.ross.ea:81} (ERT)
we have classified
both, three-parton and four-parton transition probabilities, according to 
their color factors. It is clear that the cancelation of IR divergences 
between
three-parton and four-parton processes can only occur inside groups of
diagrams with the same color factor. The cancelation of IR
divergences can be seen more clearly by representing the different
amplitudes as the different cuts one can perform in the three-loop bubble
diagrams contributing to the $Z$-boson selfenergy. 
After summing up the three-parton and four-parton contributions to the
three-jet decay width of the $Z$-boson we obtain the functions
$H_{V(A)}$ in \eq{eq:gamma3jets} at order
$\as$. 
Since a large part of the calculation has been done numerically, 
it is important to have some checks of it. We have performed the following 
tests:
i) We have checked 
our four-parton probabilities in the massless limit
against the amplitudes presented in ref.~\cite{ellis.ross.ea:81}.
The three-parton amplitudes for massive quarks cannot be compared directly 
with the corresponding massless result 
as they have different structure of IR singularities.
ii) The four-parton transition amplitudes have also been checked in the 
case of massive quarks by comparing their contribution 
to four-jet processes with the known results \cite{ballestrero.maina.ea:92}.
iii) To check the performance of the numerical programs 
we have applied our method to the massless amplitudes of ERT and 
obtained the known results for the functions $A\el{1}$. 
iv) We have checked, independently for each of the groups of diagrams
with different color factors, that the final result obtained with massive 
quarks reduces to the massless result in the limit of very small masses.

The last test is the main check of our calculation.
We have calculated the functions
$H_{V(A)}$ for several values of
$\rb$, in the range $M_b\sim 1 - 5~GeV$, and then we have extrapolated 
the results for $\rb \rightarrow 0$. 
In that limit we reproduce the values for the function
$A\el{1}$ in the different algorithms considered and the different
groups of diagrams.
This check is not trivial at all
since the structure of IR divergences for massive quarks
is quite different from the case of massless quarks: for massive quarks
collinear divergences are regulated by the quark mass, and therefore some of 
the poles in $\epsilon$ that appear in the massless case are softened by
$\log \rb$. 

Combining
\eq{eq:r3bd_def}, 
\eq{eq:gamma3jets} and \eq{eq:hexpansion} 
and using the
known expression for $\Gamma^b$ 
\cite{bilenky.rodrigo.ea:95,djouadi.kuhn.ea:90}
we write $R^{bd}_3$ as the following expansion in $\as$ 
\begin{equation}
R^{bd}_3 = 1+ \rb \left(b_0+\frac{\as}{\pi} b_1\right)~,
\label{eq:r3resultpole}
\end{equation}
where the functions $b_0$ and $b_1$ are an average of the
vector and axial-vector parts, weighted by
$c_V= g_V^2/(g_V^2+g_A^2)$ and
$c_A= g_A^2/(g_V^2+g_A^2)$ respectively. They can be written in terms of
the different functions introduced before, \eq{eq:hexpansion},
\cite{bilenky.rodrigo.ea:95,rodrigo:96} and also depend on $\yc$ and $\rb$.

It is important to note that because the particular normalization we have
used in the definition of $R^{bd}_3$, which is manifested in the final 
dependence
on $c_V$ and $c_A$, most of the electroweak corrections cancel. Those are
about 1\% \cite{bernabeu.pich.ea:91}
in total rates, while in $R^{bd}_3$ are below 0.05\%. Therefore,
for our estimates it is enough to consider tree-level values of $g_V$ and
$g_A$. The same argument applies for the passage from decay widths to
cross sections. Contributions from photon exchange are small at LEP and
can be absorbed in a redefinition of $g_V^2$ and $g_A^2$ 
\cite{jersak.laermann.ea:82}.
They will add a small correction to our observable.

Although intermediate calculations have been performed using the 
pole mass, we can also re-express our results in terms of the running
quark mass by using the known perturbative expression
$
M_b^2 = \bar{m}_b^2(\mu) [1 + 2\as(\mu)/\pi\ 
( 4/3 - \log(m_b^2/\mu^2) ) ]~.
$
The connection between pole and running masses is known up to
order $\as^2$, however consistency of our pure perturbative 
calculation requires we use only the expression above. We obtain
\begin{equation}
R^{bd}_3 = 1+ \bar{r}_b(\mu) \left(
b_0+\frac{\as(\mu)}{\pi} 
\left[\bar{b}_1-2 b_0 \log \frac{m_Z^2}{\mu^2}\right]
\right)~,
\label{eq:r3resultrunning}
\end{equation}
where $\bar{r}_b(\mu)=\bar{m}^2_b(\mu)/m_Z^2$ and
\begin{equation}
\bar{b}_1 = b_1+b_0\left[8/3-2\log(\rb)\right]~.
\label{eq:b1bar}
\end{equation}
$\bar{r}_b(\mu)$ can be expressed in terms of the running mass of
the $b$-quark at $\mu = m_Z$ by using the renormalization group. At the
order we are working
$
\bar{r}_b(\mu)=\bar{r}_b(m_Z)\left(
\alpha_s(m_Z)/\alpha_s(\mu)\right)^{-4\gamma_0/\beta_0}
$
with
$
\alpha_s(\mu) = \alpha_s(m_Z)/(1+\alpha_s(m_Z)\beta_0 t) 
\label{eq:mbrunning}
$
and $t=\log(\mu^2/m^2_Z)/(4\pi)$, $\beta_0 =11-2N_f/3$, $N_f=5$ 
and $\gamma_0=2$.

At the perturbative level \eq{eq:r3resultpole} and \eq{eq:r3resultrunning}
are equivalent. However, they neglect different higher order terms
and lead to different answers. Since the experiment is performed at
high energies (the relevant scales are $m_Z$ and $m_Z \sqrt{\yc}$) one
would think that the expression in terms of the running mass is more
appropriate because the running mass is a true short distance parameter,
while the pole mass contains in it all the complicated physics at
scales $\mu \sim M_b$. Moreover, by using the expression in terms of the
running mass we can vary the scale in order to estimate the error
due to the neglect of higher order corrections.
In any case, if one  would use in \eq{eq:r3resultrunning}  
scales as low as $\mu = 5~GeV$,
one would get something closer to the pole mass result. 

Although we have studied the observable \eq{eq:r3bd_def} for
the four jet-clustering algorithms 
discussed in 
\cite{bilenky.rodrigo.ea:95,bethke.kunszt.ea:92,rodrigo:96},
in the following we concentrate only on the DURHAM algorithm 
\cite{brown.stirling:92*b},
which gives smaller radiative
corrections and was the one used by the
DELPHI collaboration in its analysis.

The function $b_0$ gives the mass corrections at the leading order. 
As shown
in \cite{bilenky.rodrigo.ea:95} it depends very mildly on 
the quark mass
in the region of interest ( $M_b \sim 3 - 5~GeV$). Therefore it is appropriate
to present our results for $b_0$  as a fit 
in only $\yc$: $b_0 = \sum_{n=0}^2 k_0^{(n)}$ log${}^n y_c$.
For the DURHAM algorithm,
in the range $0.01 < y_c < 0.10$ and $3~GeV < M_b < 5~GeV$, using
$s_W^2 = 0.2315$, we obtain
$k_0^{(0)}=-10.521\;$,  $k_0^{(1)}=-4.4352 \;$, $k_0^{(2)}=-1.6629\;$.

The function $\bar{b}_1$~\cite{rodrigo.santamaria.ea:97*a}
gives the NLO massive corrections to $R^{bd}_3$. 
It is important to note that $\bar{b}_1$
contains significant logarithmic corrections depending on the quark mass. 
We take them into account by using the form
$\bar{b}_1= k_1^{(0)} + k_1^{(1)}\log(\yc) + k_m^{(0)} \log (\rb)$ 
in the fit.
The coefficients we obtain, for the DURHAM scheme and ranges
for $\yc$ and $\rb$ mentioned above, are:
$k_1^{(0)}=297.92\;$,  $k_1^{(1)}=59.358 \;$, $k_m^{(0)}=46.238\;$.
\mafigura{5.cm}{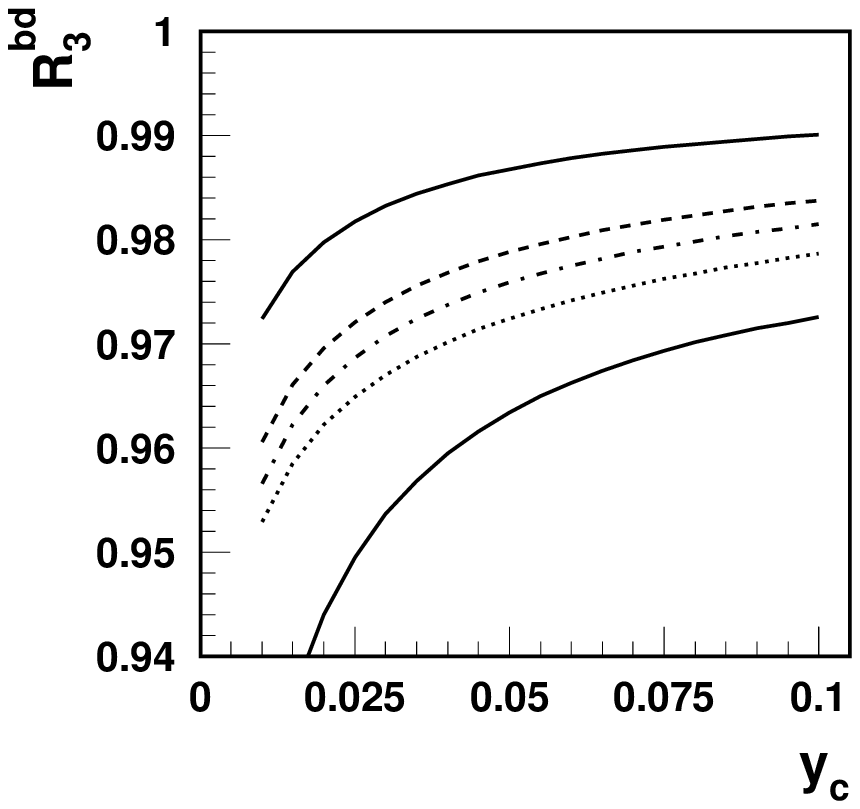}{
NLO results for $R_3^{bd}$ (DURHAM)
for $\mu=m_Z$ (dashed), 
$\mu=30~GeV$ (dashed-dotted) 
and $\mu= 10~GeV$ (dotted) for $\bar{m}_b(m_Z)= 3~GeV$ and 
$\alpha_s(m_Z)=0.118$. For comparison we also plot the LO results for
$M_b = 5~GeV$ (lower solid line) and $\bar{m}_b(m_Z) = 3~GeV$ 
(upper solid line)
}{fig:r3}

In fig.~\ref{fig:r3} we present $R_3^{bd}$
for $\mu=m_Z$ (dashed), $\mu=30~GeV$ (dashed-dotted) and $\mu= 10~GeV$
(dotted) for $\bar{m}_b(m_Z)= 3~GeV$ and $\alpha_s(m_Z)=0.118$.
For comparison we also present the LO results for the quark mass equal to
$5~GeV$ (lower solid line) which is, roughly,
the value of the pole mass obtained at low energies and
$3~GeV$ (upper solid line)
which is, roughly, the value one obtains for the running mass at the $m_Z$
scale by using the renormalization group. 
Note that choosing
a low value for $\mu$ makes the result closer to the LO result written
in terms of the pole mass, while choosing a large $\mu$ makes the result
approach to the LO result written in terms of the running mass at the
$m_Z$ scale.

\section{$\bar{m}_b(m_Z)$ from LEP data}

If $R_3^{bd}$ is measured to good accuracy one could use
\eq{eq:r3resultrunning} and the relationship between
$\bar{m}_b(\mu)$ and $\bar{m}_b(m_Z)$
to extract $\bar{m}_b(m_Z)$. However, 
the extracted
result will depend on the scale $\mu$.
For illustration, in fig.~\ref{fig:mbbar} we show 
the value for $\bar{m}_b(m_Z)$ which
one would obtain from
$R^{bd}_{3\, exp}=0.96  ~(\yc = 0.02)$
as a function of the scale $\mu$.
Although one would naturally think that the scale $\mu$ has to be
taken of the order $\sim m_Z/3$
if the energy is equally distributed among the three jets,
strictly speaking the precise value of $\mu$ is undefined.
The spread of the result due to the variation of the scale
in an appropriate range gives
an estimate of the  uncertainty due
to higher order corrections. From fig.~\ref{fig:mbbar}
we see that if we vary $\mu$ in the range
$m_Z/10 - m_Z$ the uncertainty in the determination
of $\bar{m}_b(m_Z)$ would be of about $0.20~GeV$ for the DURHAM algorithm.
In the same range and for the JADE and EM algorithms the obtained 
error would be~\cite{rodrigo:96} of about $0.25~GeV$,
while for the E algorithm we would obtain an error
bigger than $0.50~GeV$.

\mafigura{5.cm}{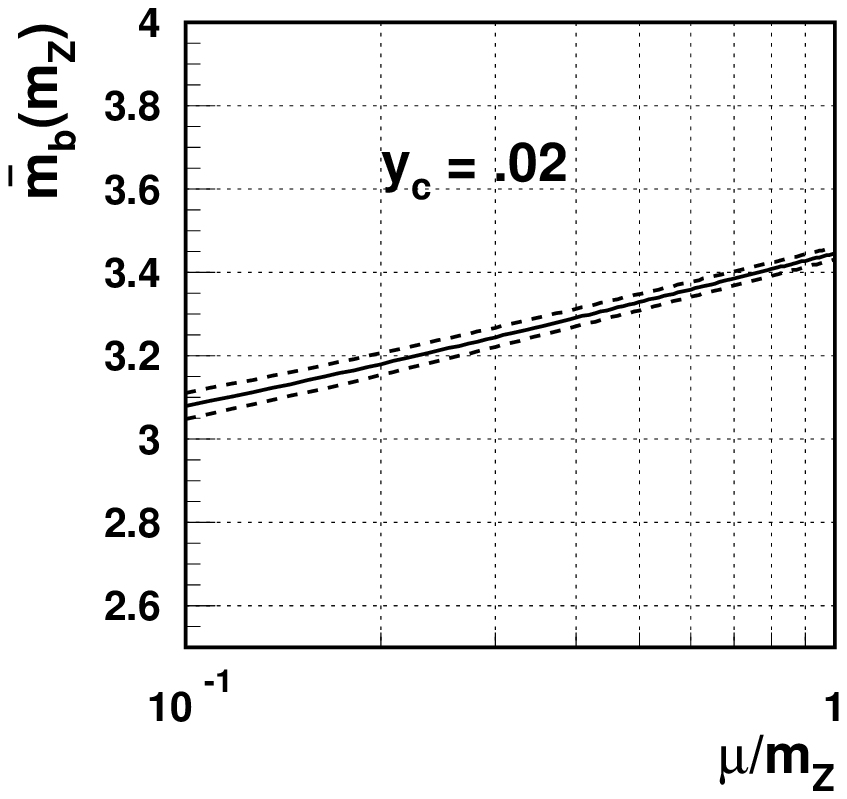}{
Extracted value of $\bar{m}_b(m_Z)$ if $R^{bd}_{3\, exp}=0.96$ as
a function of the scale $\mu$. We take $\alpha_s(m_Z)=0.118$ (solid)
and $\Delta \as = 0.003$ (dashed).
}{fig:mbbar}

The DELPHI collaboration has used a slightly different
approach to extract the value of $\bar{m}_b(m_Z)$
from $R_3^{bd}$~\cite{marti.fuster.ea:97}.
Using \eq{eq:r3resultpole}
written in terms of the pole mass, $M_b$, and, exploiting 
the perturbative relation
between the pole mass and the running mass, they obtained the value
of $\bar{m}_b(m_Z)$ from $M_b$ extracted from the experimentally 
measured $R_3^{bl}$.
The difference between $\bar{m}_b(m_Z)$ obtained in this way and
the running mass obtained directly using~\eq{eq:r3resultrunning},
which is due to the different treatments 
of the higher order terms, was considered as a theoretical
uncertainty. Accounting in addition for the uncertainty due to the 
variation 
of the scale $\mu$ they get a more conservative estimate for the 
theoretical error, $0.27~GeV$, for the DURHAM algorithm.
We would like to note, however, that the central value for 
$\bar{m}_b(m_Z)$
reported in~\cite{marti.fuster.ea:97} is fully compatible with
the one obtained by using only~\eq{eq:r3resultrunning}. 
The comparison of the two
methods gives a check of consistency.
Furthermore, the mass of the bottom-quark, $\bar{m}_b(m_Z)$, measured from the 
three-jet decay of the Z-boson~\cite{marti.fuster.ea:97}
is also fully compatible with the value 
obtained from low energy determinations~\cite{jamin.pich:97} 
after using the renormalization group. This provides, for the first time,
a nice check of the running of a quark mass.

\vskip 5mm

\noindent{\bf Acknowledgements.} 
We are indebted with S. Cabrera, 
J. Fuster and S. Mart\'{\i} for a very 
enjoyable collaboration. G.R. acknowledges O.~Biebel
for very useful comments 
and S. Narison for the very kind atmosphere created 
at Montpellier.



\end{document}